\documentclass[aps,pra, twocolumn]{revtex4-1}
\usepackage[utf8]{inputenc}
\usepackage{amsmath}
\usepackage{amsfonts}
\usepackage{mathtools}
\usepackage{dsfont}
\usepackage{braket}
\usepackage{tikz}
\usetikzlibrary{backgrounds,fit,decorations.pathreplacing}
\usetikzlibrary{trees}
\usepackage{qcircuit}
\usepackage[ruled,vlined]{algorithm2e}

\newtheorem{theorem}{Theorem}
\newtheorem{result}{Result}
\newtheorem{lemma}{Lemma}
\newtheorem{definition}{Definition}

\newcommand{\prlsection}[1]{ \emph{#1}.--}

\begin{document}

\title{Quantum theory cannot violate a causal inequality}
\author{Tom Purves}\email{tom.purves@bristol.ac.uk} \affiliation{H.H. Wills Physics Laboratory, University of Bristol, Tyndall Avenue, Bristol, BS8 1TL, U.K.}

\author{Anthony J. Short}\email{tony.short@bristol.ac.uk} \affiliation{H.H. Wills Physics Laboratory, University of Bristol, Tyndall Avenue, Bristol, BS8 1TL, U.K.}


\begin{abstract}
  Within quantum theory, we can create superpositions of different causal orders of events, and observe interference between them. This raises the question of whether quantum theory can produce results that would be impossible to replicate with any classical causal model, thereby violating a causal inequality. This would be a temporal analogue of Bell inequality violation, which proves that no local hidden variable model can replicate quantum results. However, unlike the case of non-locality, we show that quantum experiments \emph{can} be simulated by a classical causal model, and therefore cannot violate a causal inequality.
\end{abstract}

\maketitle

\prlsection{Introduction} A fascinating aspect of quantum theory that has been investigated recently is the possibility for the causal order of events to be placed into superposition \cite{ Chiribella2013, Chiribella2012,Branciard2016, Zych2019, Barrett2020}, leading to `causal indefiniteness' about the order with which events have taken place. This phenomenon has been tested experimentally \cite{Procopio2015, White2018, Goswami2020}, and can be exploited to gain advantages within quantum theory. For example, setups based on the quantum switch \cite{Chiribella2013} can help to determine whether unknown unitaries commute or anticommute \cite{Chiribella2012}. An interesting question is whether quantum theory can generate results which could not be simulated by any classical causal model. Such results would violate a 
 Causal Inequality \cite{Oreshkov2012, Brukner2014, Branciard2015, Abbott2016}. These are the temporal analogues of Bell Inequalities \cite{Bell1964}, and the violation of such an inequality in nature would call into question the elementary properties that scientists regularly invoke when talking about cause and effect relationships. 

In this paper we focus on the relationship between the type of causal indefiniteness present in quantum theory and the type needed to violate causal inequalities. We show that despite allowing causally indefinite processes, the correlations generated by quantum theory can be simulated by a classical causal model. This means that quantum theory cannot violate causal inequalities, and hence cannot yield an advantage over classical causal processes for tasks defined in a theory-independent way (such as `guess your neighbour's input' \cite{Almeida2010, Branciard2015}). Previous works in this direction have shown that particular switch-type scenarios cannot violate causal inequalities \cite{Arajo2015}, and that causal order cannot be placed in a pure superposition \cite{Costa2020, Yokojima2021}. It has also been shown that causal inequality violations are possible when we condition on measurement outcomes of one party \cite{Milz2020}. However, our results imply that such violations are not possible for general quantum setups without conditioning. 

Indefinite causal structure is often studied via process matrices \cite{Oreshkov2012}, which assume that local laboratories obey standard quantum theory, but allow any connections between them consistent with this. This may include processes which are not achievable in standard quantum theory, or in nature more generally. Here we focus on what is possible in standard quantum theory, using quantum control of different parties' operations to generate superpositions of causal order, in a similar way to \cite{Colnaghi2012,Araujo2014}. As process matrices can yield causal inequality violation, a corollary of our result is that all process matrices cannot be implemented in standard quantum theory.


\prlsection{Results}\label{sec::1} Before considering quantum processes, we first define causal processes, which are those which could be realised classically by a set of parties in separate laboratories passing systems between them \cite{pearl, Barrett2020}. 

\begin{figure}[h]
    \centering
    \includegraphics[width=0.46\textwidth]{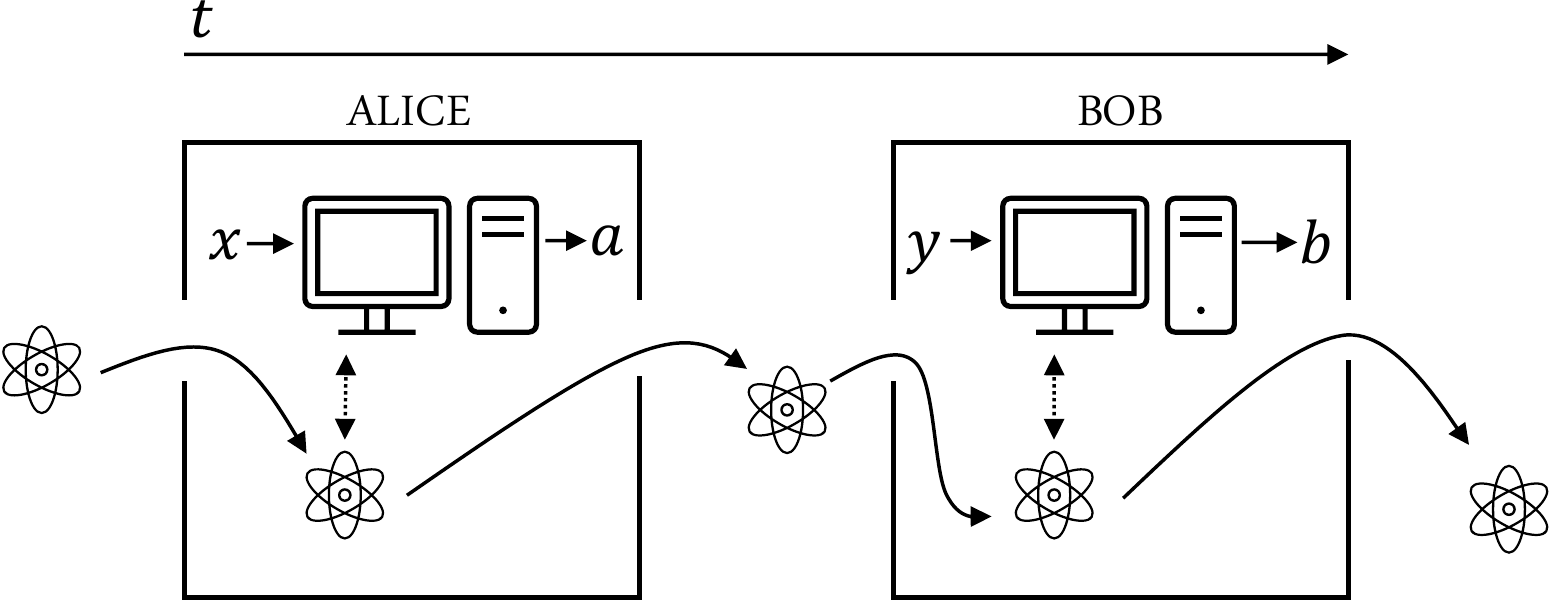}
    \caption{
    An example of a causal process in which Alice goes before Bob. Note that the system which is passed from Alice's to Bob's laboratory could encode information about $a$ and $x$.}
    \label{fig:causalcorrilation}
\end{figure}

First consider two parties, Alice and Bob, with measurement settings $x$ and $y$ and measurement results $a$ and $b$ respectively. During the experiment, depicted in figure \ref{fig:causalcorrilation}, each party sees a system enter their laboratory exactly once, performs a measurement on it with their corresponding measurement setting (which may also modify the system), and records their result. They then pass the system out of their laboratory. Apart from the systems entering and leaving their laboratories, the two parties cannot communicate with each other, but the system leaving one laboratory may be later sent into the other. Alice and Bob's joint measurement results can be described by a conditional probability distribution $p(ab|xy)$. However, not all such probability distributions can be achieved by a causal process.

The most general causal process in this case would be to first choose randomly whether Alice or Bob would go first (with probabilities $p(\textrm{Alice first})$ or $p(\textrm{Bob first})$). If Alice goes first, then her measurement result can depend on her measurement setting but not on Bob's, who hasn't acted yet, so is given by $p(a|x)$. She can then encode her measurement setting and result in the system and pass it out of her laboratory. This system then enters Bob's laboratory, where his measurement result can depend on all of the other variables, given by $p(b|a,x,y)$. Considering the other causal order in which Bob goes first in the same way, we obtain \cite{Branciard2015}
\begin{align} \label{eq:twopartycausal}
p^{\mathrm{causal}}(ab|xy) = &p(\textrm{Alice first}) p(a|x) p(b|a,x,y) \nonumber \\
&+ 
p(\textrm{Bob first}) p(b|y) p(a|b,x,y)
\end{align}

For the multiparty generalisation \cite{Oreshkov2016multi, Abbott2016},  observe that the above causal probability contains two types of terms. The first, such as $p(\textrm{Alice first})$, determines the order in which the parties act, and the second, such as  $p(a|x)$ or $p(b|a,x,y)$, determines the outcome probabilities of their measurements, constrained by their causal order. We now extend these `who is next?' and `what did they see?' type probabilities to an arbitrary number of parties. We use $l_k$ to denote the $k{\textrm{th}}$ party that receives the system (or equivalently, the $k{\textrm{th}}$ laboratory the system enters), and denote the probability for this to occur by $p_k(l_k|H_{k-1})$. The conditional on $H_{k-1}$ represents the history (including all previous parties that have measured, and their inputs and outputs) for it should be permitted for parties in the causal past of $l_k$ to affect who is the next party to act. As a simple example of this, consider a tripartite experiment, with Alice, Bob and Charlie participating. If Charlie comes first,  the system could be passed to Alice or Bob next, based on the outcome of his measurement. Here, $p_2(\text{Alice next}|\text{Charlie got outcome = }1)$ may not be equal to $p_2(\text{Alice next}|\text{Charlie got outcome = }0)$. Scenarios of this form this are what $p_k(l_k|H_{k-1})$ accounts for. The probability for $l_k$ to obtain given results may also depend on this history (but, importantly, not on the causal future), and of course on the measurement setting, denoted $x_{l_k}$. We write this probability as $p_k(a_{l_k}|H_{k-1},x_{l_k})$. A causal model is then the summation over all available parties at all stages of the measurement procedure, under the assumption  that each party  only  acts once in the entire procedure. 

\begin{definition}\label{def::model}
A causal probabilistic model can be written as
\begin{align} \label{eq:classical}
    p^{\mathrm{causal}}(\vec{a}|\vec{x})=
    \sum_{l_1 \notin \mathcal{L}_0} ... \sum_{l_N \notin \mathcal{L}_N-1} &p_1(l_1|H_0)p_1(a_{l_1}|H_0,x_{l_1})... \nonumber\\
   \quad ...
    p_N(l_N|H_{N-1}&)p_N(a_{l_N}|H_{N-1},x_{l_N})
\end{align}
where the $p_k(l_k|H_{k-1})$ terms represent probabilities for party $l_k$ to act at stage $k$ of the causal order, and $p_k(a_{l_k}|H_{k-1},x_{l_k})$ terms represent probabilities for party $l_k$, who has acted at stage $k$ of the causal order to obtain measurement result $a_{l_k}$. Both of the above probabilities are conditional on a history, $H_{k-1}$, which contains all of the information about previous inputs, outputs and party order. In particular, the history $H_{k}=(h_1,...,h_k)$ is the ordered list of triples $h_i=(l_i,a_{l_i},x_{l_i})$. The summations are performed over all possible next parties, excluding parties who have already acted, which are stored in the unordered sets $ \mathcal{L}_k=\{l_1,...,l_k\}$. To emphasise the symmetry between the terms we include $H_0$ and  $\mathcal{L}_0$, which are defined as empty sets, as no parties have acted at that point.
\end{definition}

This definition leads to a convex polytope of causal probability distributions $p^{\mathrm{causal}}(\vec{a}|\vec{x})$. Note that although the notation differs, this generates the same set of probabilities as was previously defined in \cite{Abbott2016, Oreshkov2016multi}. Linear constraints on these probabilities which are satisfied by all  $p^{\mathrm{causal}}(\vec{a}|\vec{x})$ but which could be violated by some arbitrary probability distribution  $p(\vec{a}|\vec{x})$ are known as `causal inequalities', and are a temporal analogue of the Bell inequalities which have been widely studied in the context of quantum non-locality. By definition, any $p^{causal}(a,b|x,y)$ cannot violate a causal inequality. A violation of a causal inequality, by observation in experiment or by calculation in theory, proves that those experimental results or predictions do not have a causal explanation of the type defined above.



\prlsection{Quantum Processes}
A general representation of quantum theory is provided by the quantum circuit model. However, if we construct a circuit with the parties' actions at fixed locations, then there is no causal indefiniteness and a causal inequality cannot be violated \footnote{We could space out the circuit such that there is at most one lab at each time-step, and then pass the full quantum state between the labs as a classical hidden variable which would allow us to recover the same correlations in a causal way}. Even to capture all classical causal processes, we need to be able to alter when different parties act. This can be achieved in the circuit model by representing the parties' actions by controlled quantum gates. Such gates could  be constructed within standard quantum theory (e.g. by sending a system into the lab when the control is in the appropriate state and not otherwise), and  are effectively what has been used in experiments probing quantum causality \cite{White2018}. Quantum circuits involving controlled lab gates appear sufficient to represent any processes achievable within standard quantum theory.   

For simplicity, we consider a setup involving a single quantum control which can trigger any of the labs. However this is equivalent to considering any quantum circuit which can be constructed from any number of individual controlled lab operations and other unitary gates (see the appendix for more details). 

The key idea is to consider $N$ parties, each of whom will interact with a quantum system exactly once, but in an order that is controlled coherently via the quantum control. We allow arbitrary unitary transformations of the system and control between each party's action, so that the ordering of later parties can be modified by earlier actions.  

\begin{figure*}
    \centering
    \includegraphics[width=\textwidth]{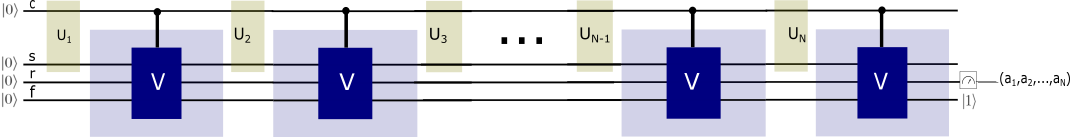}
    \caption{Illustration of the quantum protocol. The system interacts with the different parties via a sequence of controlled entangling unitaries. Quantum control of causal order is achieved by a series of unitaries $U_n$ on the control and system wires.}
    \label{fig:pictureofproto}
\end{figure*}

To allow the maximum possible interference, and avoid `collapses' which would prevent  interference between different causal orders, we model each party's measurement as a unitary interaction between the system and a local measurement register. This corresponds to the case in which there is no record in the measuring device of the time at which the measurement was performed. At the end of the experiment, all parties read off their measurement results from their local measurement registers (which can be modelled by a standard projective measurement). 

Each party also has a `flag' which keeps track of how many times they have interacted with the system. At the end of the protocol we require that each party has interacted with the system exactly once. 

Formally, the Hilbert space can be decomposed into the following components \begin{itemize}
\item An arbitrary quantum system $\mathcal{H}_s$, which is passed between parties. 

\item A quantum control $\mathcal{H}_c$ which has dimension $N+1$.
The  basis states $\ket{1}\ldots \ket{N}$ denote which party will measure next, while $\ket{0}$ is treated as a `do nothing' command. By considering superpositions of these basis states, we can superpose different causal orders.

\item A result register $\mathcal{H}_{r_i}$ for each of the $N$ parties. The different results are represented by orthonormal basis states $\ket{a_i}$ with $a_i \in \mathcal{A}_i$, leading to the result register having dimension $|\mathcal{A}_i|$. We choose one of these states as a starting state for the results register and denote it by $\ket{0}_{r_i}$.

\item A `flag' $\mathcal{H}_{f_i}$ for each of the $N$ parties, indicating how many times they have interacted with the system. For simplicity, we take each of these to be infinite dimensional, with basis states labelled by the integers. When the party interacts with the system the value of the flag is unitarily raised by the operator $\Gamma=\sum_{n}\ket{n+1}_{f_i}\bra{n}_{f_i}$. Each flag starts in the $\ket{0}_{f_i}$ state, and at the end of the protocol, we require them all to be in the $\ket{1}_{f_i}$ state.  

\end{itemize} 

Note that we do not include separate local quantum ancillas for the parties, as these can always be incorporated in $\mathcal{H}_s$. We  denote the combined result and flag spaces by  $\mathcal{H}_r = \bigotimes{\mathcal{H}_{r_i}}$ and  $\mathcal{H}_f = \bigotimes{\mathcal{H}_{f_i}}$ respectively. 

We consider quantum protocols as follows. Firstly, the initial state
\begin{align}
    \ket{0}=\ket{0}_s\ket{0}_c\ket{0}_r^{N}\ket{0}_f^{N}\in \mathcal{H}_s\otimes \mathcal{H}_c \otimes \mathcal{H}_r \otimes \mathcal{H}_f.
\end{align}
is prepared, and each party $l$ either chooses or is distributed their individual classical measurement setting $x_l$.

The protocol then consists of $T$  time-steps, each of which is composed of two operations. Firstly, an arbitrary unitary transformation $U_t$ is applied to the system and control, which can depend on the time $t$. Secondly,  a fixed controlled lab-activation unitary $V$ is applied, which activates whichever party is specified by the control. This is given by 
\begin{equation} \label{eq::veqn}
V=\ket{0}\bra{0}_c \otimes I + \sum_{l=1}^{N} \ket{l}\bra{l}_c \otimes V_{s,r_{l}}(x_{l}) \otimes \Gamma_{f_{l}} \otimes  I 
\end{equation}
where the identities are over all remaining subsystems. $V_{s,r_{l}}(x_{l})$ is a unitary which implements the measurement of party $l$  on the system specified by the measurement setting $x_l$, and stores the result in the register $r_l$. For example, two different values of $x_l$ could correspond to party $l$ measuring the system in either the computational or the Fourier basis. Note that by incorporating ancillas within the system, any local quantum measurement (i.e a POVM) is realisable within this paradigm. Ancillas can also be used to generate arbitrary mixed states if required (via purification).

The unitary operator $\Gamma_{f_{l}}$ raises the flag system of the party making their measurement.  At the end of the protocol, we require that the flags are in the state $\ket{1}_f^{N}$ (i.e. that each party has measured the system once). This places constraints on the possible protocols which can be constructed. Note that each party does not have access to an operation which resets the flag, aside from the initialisation operation at the start of the protocol. They therefore  always `remember' if they have made a measurement or not. Also, we do not allow circuits involving the controlled inverse of a party's action (which would lower their flag and erase their memory), as this would enlarge the set of causal possibilities even classically.

The total unitary for the protocol is  given by 
\begin{equation} 
\mathcal{U}=VU_TVU_{T-1}...VU_{1}.
\end{equation} 
At the end of the protocol, each party performs a projective measurement on their results register to obtain their final result \footnote{Note that from a many-worlds perspective \cite{Everett} such an additional step would not be necessary. However, we include it here to maintain connection with standard quantum theory and give an explicit formula for the outcome probabilities}. The output probability distribution of the quantum protocol is therefore given by 
\begin{align} \label{eq:quantumdist} 
    p^{\mathrm{quantum}}(\vec{a}|\vec{x})=|(\ket{\vec{a}}\bra{\vec{a}}_r \otimes I)\mathcal{U}\ket{0}|^2. 
\end{align}
The full protocol is illustrated as a quantum circuit in figure \ref{fig:pictureofproto}.

The main result of this paper is that any probability distribution  which can be generated within quantum theory, as described above, can also be obtained via a classical causal process. 

\begin{theorem}
Any quantum probability distribution $p^{\mathrm{quantum}}(\vec{a}|\vec{x})$ can be exactly replicated by a classically causal process $p^{\mathrm{causal}}(\vec{a}|\vec{x})$. Hence quantum theory cannot violate a causal inequality.
\end{theorem}

In particular, we now show how to construct an explicit classical causal process which replicates the results of any quantum protocol, together with a sketch of the proof of Theorem 1. The full proof of the theorem can be found in the appendix. 

We first define notation for describing states at each stage of the quantum protocol, and then show how to use these to construct the probabilities in the corresponding classical model. 

\begin{definition}
The (un-normalised) state  with a History $H_{k-1}$, at a time  $t$, with the control set to trigger the action of party $l_k$ is given by 
\begin{align}
    \ket{\psi_{(l_k,t,H_{k-1})}}= (\ket{l_k}\bra{l_k}_c \!\otimes\! \pi^{H_{k-1}}_{rf} \!\otimes\! I_s)U_t V U_{t-1}...V U_1 \ket{0}.
\end{align}
 The projector onto the result and flag spaces is given by
$
     \pi^{H_{k-1}}_{rf} = \bigotimes_{i=1}^{N}\left(\pi^{H_{k-1}}_{r_{i}f_{i}}  \right),
$
where
\begin{align}
    \pi^{H_{k-1}}_{r_{i}f_{i}} = \begin{cases} 
    \ket{a_{i}}\bra{a_{i}}_{r_{i}} \otimes  \ket{1}\bra{1}_{f_{i}} & \text{ if } (i, a_i, x_i) \in H_{k-1},\\
   I_{r_{i}} \otimes \ket{0}\bra{0}_{f_{i}} & \text{ otherwise }.
    \end{cases} 
\end{align}
\end{definition}
This notation describes states which are about to be measured by the parties (i.e., a $V$ type operator is about to act on them). We also set up some notation for states which have just been measured, in a similar fashion.

\begin{definition}
The (un-normalised) state with a History $H_{k}$, at a time $t$, in which party $l_k$ has just acted is given by
\begin{align}
    \ket{\phi_{(l_k,t,H_{k})}}=(\ket{a_{l_k}}\bra{a_{l_k}}_{r_k} \otimes I)V\ket{\psi_{(l_k,t,H_{k-1})}}.
\end{align}

\end{definition}
With these definitions, we can associate the states in this quantum process with the probabilities in our classical causal model.
\begin{definition}
The probability for party $l_k$ to act next, given a history $H_{k-1}$ is given by:
\begin{align} \label{eq:probnextm}
    p_k(l_k|H_{k-1})=\frac{\sum_{t_k=1}^{T}|\ket{\psi_{(l_k,t_k,H_{k-1})}}|^2}{\sum_{l_k' \notin \mathcal{L}_{k-1}}\sum_{t_k'=1}^{T}|\ket{\psi_{(l_k',t_k',H_{k-1})}}|^2}.
\end{align}
\end{definition}

We have summed over time \cite{Note2}, because it is possible within the quantum paradigm to conduct the $k^{th}$ measurement at different times according to a background clock (which we note the labs have no access to). Note that states at different times combine incoherently, but different sequences leading to the same set of historical measurement results combine coherently inside $\ket{\psi_{(l_k,t_k,H_{k-1})}}$.

The form of  equation \eqref{eq:probnextm} makes it a valid probability distribution, as it is non-negative, and obeys the correct normalisation that $\sum_{l_k \notin \mathcal{L}_{k-1}}p_k(l_k|H_{k-1})=1$. Also note that it depends on only those input variables $x_i$ which appear in the history $H_{k-1}$.

Next, we specify similar probabilities for seeing measurement results based on a given history.
\begin{definition}
The probability for party $l_k$ to obtain the measurement result $a_{l_k}$, given a history $H_{k-1}$, and an input variable $x_{l_k}$ is given by:
\begin{align}\label{eq:probresultm}
    p_k(a_{l_k}|H_{k-1},x_{l_k})=\frac{\sum_{t_k=1}^{T}|\ket{\phi_{(l_k,t_k,H_k)}}|^2}{\sum_{a'_{l_k} \in \mathcal{A}_{l_k}} \sum_{t_k'=1}^{T}|\ket{\phi_{(l_k,t_k',H_{k}')}}|^2}, 
\end{align}
where $H_k = (H_{k-1}, (l_k, a_{l_k}, x_{l_k}))$ and $H_k' = (H_{k-1}, (l_k, a'_{l_k}, x_{l_k}))$
\end{definition}

This is again a valid probability distribution, since $\sum_{a_{l_k} \in \mathcal{A}_{l_k}} p_k(a_{l_k}|H_{k-1},x_{l_k})=1$. In the numerator, we have simply taken sum of the modulus squared of all of the states which have the correct historical results, the control in the correct state, and the results register containing the result we want to calculate the probability for. 

To prove Theorem 1, We begin by inserting $ p_k(l_k|H_{k-1})$ (from \eqref{eq:probnextm}) and $p_k(a_{l_k}|H_{k-1},x_{l_k})$ (from \eqref{eq:probresultm})  
 into the definition of a causal model \eqref{def::model}. We are then able to straightforwardly cancel the numerator of the `who is next?' type probabilities with the denominator of the `what did they see?' probabilities for the probabilities evaluated \textit{at the same stage of the causal order}. Next, we show that a sum over the last party to measure in the numerator at one stage of the causal order, cancels with the denominator \textit{at the next stage of the causal order}. We then make the observation that for the first stage of the causal order, the denominator of $p_1(l_1 | H_0)$ is equal to one (which corresponds to the fact that someone must measure first in the quantum circuit). Finally, we note that the numerator of the final term, summed over all parties, represents exactly the  probabilities $p^{\mathrm{quantum}}(\vec{a}|\vec{x}) $ arising from the quantum protocol. This allows us to simulate the results of the quantum protocol via the classically causal model given in \eqref{def::model}. Given that it can be replicated by a causal model, it follows that quantum theory cannot violate a causal inequality.

In the appendix, we give an example of how these results can be applied in practice, based on the quantum switch \cite{Chiribella2013}. This involves the causal order of two parties becoming entangled with the control. A third party then performs a measurement which leads to interference between the two causal orders.  It has already been shown that this simple setup cannot be used to violate a causal inequality \cite{Arajo2015, Oreshkov2012}. However, it is instructive to see how it fits into our framework. Despite the quantum setup including interference, our results give an explicit classical causal process which generates the same behaviour (i.e. the same $p(a,b,c|x,y,z)$).  

\prlsection{Conclusions} \label{sec::5} By using a quantum control to determine when different parties measure, and treating these measurements as coherent unitary processes, quantum theory allows us to generate superpositions of causal orders and to observe interference between them. At the level of the theory, such processes do not arise from a single causal order, or even a mixture of orders. However, we have shown that the probabilities $p(\vec{a}|\vec{x})$ generated by any quantum protocol \emph{can} be simulated by a classical causal process. This means that quantum theory cannot violate a causal inequality, and thus one could not convince a sceptic that nature deviates from classical notions of causality. 

This is in sharp distinction to non-locality, where not only does the theory appear non-local (e.g. via entangled states) but we can also prove that some quantum probabilities cannot be replicated by any local hidden variable model. By violating a Bell inequality we can therefore prove non-locality experimentally. 

Although our  framework is very general, one key requirement is that each party interacts with the system once (which leads to a requirement on the final flag state). This is the normal setup for causal inequalities, and allows us to assign a single input and output to each party, and to represent the experimental results via $p(\vec{a}|\vec{x})$. However, it would be interesting  to lift this assumption in future research. For example, could we obtain a violation of causality if parties are allowed to measure twice, or a variable number of times, or to forget they have measured? We also have a technical assumption that the protocol takes finite time (i.e. that it terminates after a finite number of steps). This seems physically reasonable, but it might be interesting to investigate lifting this assumption, as well as to consider extending the results to continuous time. Finally, it would be interesting to consider a network structure in the causal scenario, in the non-local case this is known to generate non-linear Bell inequalities, and sets of non convex probability polytopes \cite{Brunner2020}. Investigation of causal indefiniteness and causal inequalities in these type of scenarios might prove of general interest. 

Finally, our framework assumes standard quantum theory. If the theory changes significantly to incorporate quantum gravity we might expect new possibilities for causal inequality violation, although not necessarily \cite{Zych2019} (note that even classical general relativity allows for the existence of closed time-like curves, which appear to violate the simple classical causal models we have considered here \cite{Lobo2010, Barrett2020cyclic, Arajo2017}). We hope that the framework and tools developed here will prove helpful in discussing these interesting issues, and in highlighting differences from the standard case. 

\vspace{0.5cm}

T.Purves acknowledges support from the EPSRC.

\vspace{0.5cm} 

\emph{Note added -} Independently obtained related results using the process matrix formalism \cite{Wechs2021} appeared on the ArXiv on the same day as this paper.

\appendix
\onecolumngrid
\section{Equivalence of combined and individual controlled lab gates}

In this section we show that the framework for quantum processes in the main text is equivalent to considering any quantum circuit built up of standard unitary gates and controlled gates for individual laboratories, in terms of the probability distributions they can generate. The key ingredient is to show how to construct individual controlled lab gates from $V$, and conversely how to construct the operation $V$ from individual controlled lab gates. 

To map any circuit involving individual controlled lab gates into our framework, we first space out the gates in the circuit, so that there is only one gate per time-step (this  will increase the depth, but not affect the results). If an individual lab gate acts on only part of the system, we extend it such that it acts on the entire system, taking the action to be trivial (i.e. tensored with the identity) on any part of the system which was not initially included. We can 
then replace each individual controlled lab gate by a circuit fragment involving one use of $V$, using the approach described below. Finally we merge all unitary gates between instances of $V$ into the unitaries $U_t$. This will lead to a circuit in our framework yielding exactly the same results as the original circuit. To go in the other direction, we simply replace each instance of $V$ with its construction in terms of individual controlled lab gates.  

Figures 1 and 2 show how to construct an individual controlled lab gate using $V$. Figure 3 shows how to construct $V$ from individual controlled lab gates. 

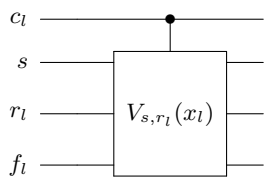
\begin{figure}[h]\
\centering
\Qcircuit @C=1.5em @R=1.2em {
&\lstick{c_l} &\qw & \ctrl{1} & \qw \\
&\lstick{s} &\qw & \multigate{2}{V_{s,r_{l}}(x_{l})} & \qw \\
&\lstick{{r_{l}}} &\qw  & \ghost{V_{s,r_{l}}(x_{l})} & \qw \\
&\lstick{{f_{l}}} &\qw  &  \ghost{V_{s,r_{l}}(x_{l})} & \qw \\
}\caption{A controlled lab gate for an individual laboratory}
\label{fig:newcircsingle}
\end{figure}

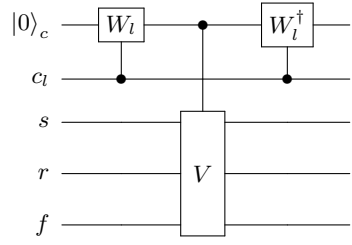
\begin{figure}[h]\
\centering
\Qcircuit @C=1.5em @R=1.2em {
&\lstick{\ket{0}_{c}} &   \gate{W_l}& \ctrl{2} & \gate{W_l^{\dagger}}& \qw\\
&\lstick{c_l} &\ctrl{-1} & \qw & \ctrl{-1} & \qw\\
&\lstick{s} &\qw & \multigate{2}{V} & \qw& \qw \\
&\lstick{r} &\qw  & \ghost{V} & \qw & \qw\\
&\lstick{f} &\qw  &  \ghost{V} & \qw & \qw\\
}\caption{An equivalent circuit to the individual controlled lab gate above, built from a single instance of $V$, where $W_l\ket{0}=\ket{l}$. Note that the individual wires may represent composite subsystems rather than individual qubits. 
\label{fig:newcircshort}}
\end{figure}

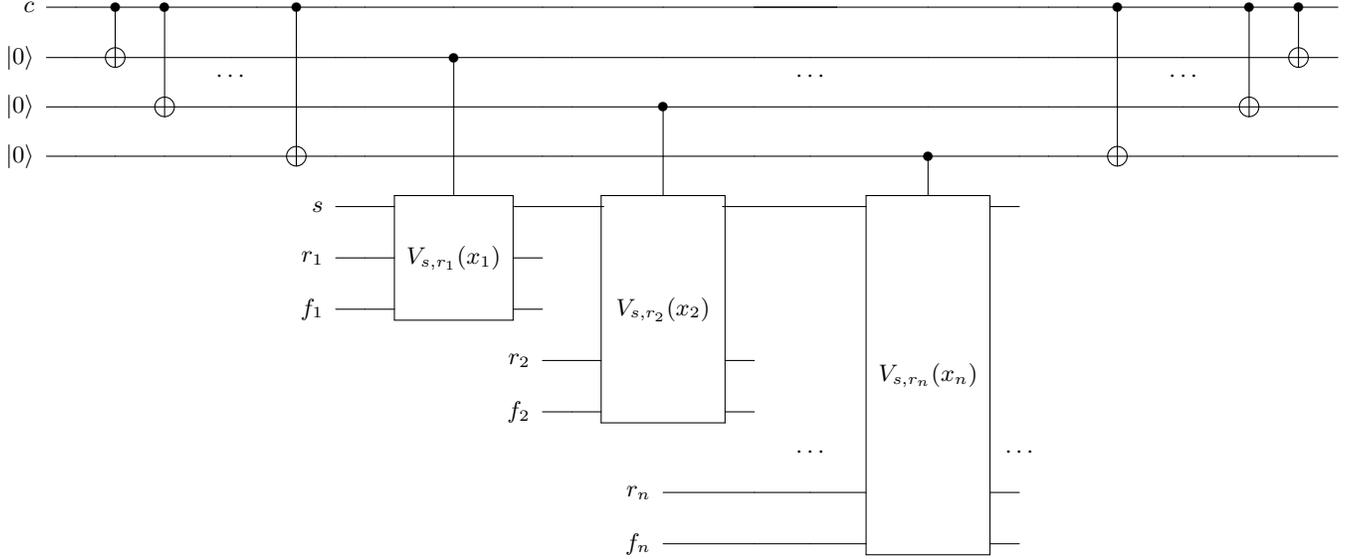
\begin{figure}[h!]\
\centering
\Qcircuit @C=1.2em @R=1.2em {
&\lstick{c}      &\qw & \ctrl{1} & \ctrl{2} & \cds{3}{\cdots} & \ctrl{3} & \qw & \qw & \qw & \qw& \qw&\qw & \qw&\cds{3}{\cdots} \qw& \qw&\qw&\qw&\qw& \ctrl{3} & \cds{3}{\cdots} & \ctrl{2} & \ctrl{1}& \qw \\
                            &  \lstick{\ket{0}}&\qw & \targ & \qw & \qw & \qw & \qw & \qw &\ctrl{3} & \qw& \qw&  \qw&\qw & \qw&\qw&\qw&\qw&\qw&\qw&\qw&\qw&\targ& \qw\\ 
& \lstick{\ket{0}}&\qw & \qw & \targ  & \qw & \qw &\qw & \qw &\qw & \qw& \qw&\ctrl{2}& \qw&\qw&\qw&\qw&\qw&\qw&\qw&\qw&\targ&\qw& \qw\\ 
                            & \lstick{\ket{0}} &\qw  & \qw & \qw & \qw &\targ&\qw & \qw &\qw & \qw&\qw& \qw & \qw& \qw &\ctrl{1}&\qw &\qw&\qw&\targ&\qw&\qw&\qw& \qw\\ 
                           & & & & & & &\lstick{s} &\qw & \multigate{2}{V_{s,r_{1}}(x_{1})} & \qw & \qw  &\multigate{4}{V_{s,r_{2}}(x_2)} & \qw &\qw & \multigate{7}{V_{s,r_{n}}(x_n)} & \qw \\
                           & & & & & & &\lstick{{r_{1}}} &\qw  & \ghost{V_{s,r_{1}}(x_{1})} &\qw  \\
                           & & & & & & &\lstick{{f_{1}}} & \qw &  \ghost{V_{s,r_{1}}(x_{1})} & \qw \\
                           & & & & & & & & & &\lstick{{r_{2}}}  & \qw& \ghost{V_{s,r_{n}}(x_{n})} & \qw\\
                           & & & & & & & & & &\lstick{{f_{2}}}  & \qw& \ghost{V_{s,r_{n}}(x_{n})} & \qw\\
                           & & & & & & & & & & & & & & \cdots & & \cdots \\
                           & & & & & & & & & & & &  \lstick{{r_{n}}} & \qw&\qw& \ghost{V_{s,r_{n}}(x_{n})} & \qw\\
                           & & & & & & & & & & & &  \lstick{{f_{n}}} & \qw&\qw& \ghost{V_{s,r_{n}}(x_{n})} & \qw\\
}\caption{V, built from individual controled lab gates. The first and last $CNOT$ gate are controlled from the state $\ket{1}_c$, the second and second from last are controlled from $\ket{2}_c$, and so on until the $n$'th and $n+1$'th $CNOT$, which are controlled from state $\ket{n}_c$.}
\label{fig:newcirclong}
\end{figure}

\newpage
\section{Proof of the Main Result} In this appendix section, we give the full proof of the main result, that the probabilities generated by a quantum protocol can be replicated by a classical causal model, and therefore cannot violate a causal inequality. We begin with recalling a few definitions from the main text, together with some convenient derived quantities. Note that we assume throughout that laboratory labels $l$ are non-zero.

\begin{definition}\label{def::modelsi}
A causal probabilistic model can be written as
\begin{align}
    p(\vec{a}|\vec{x})=\sum_{l_1 \notin \mathcal{L}_0} \sum_{l_2 \notin \mathcal{L}_1} ... \sum_{l_N \notin \mathcal{L}_N-1} p_1(l_1|H_0)p_1(a_{l_1}|H_0,x_{l_1}) p_2(l_2|H_1)p_2(a_{l_1}|H_1,x_{l_1})...
    p_N(l_N|H_{N-1})p_N(a_{l_N}|H_{N-1},x_{l_N})
\end{align}
where $p_k(l_k|H_{k-1})$ terms represent probabilities for party $l_k$ to act at stage $k$ of the causal order, and $p_k(a_{l_k}|H_{k-1},x_{l_k})$ terms represent probabilities for party $l_k$, who has acted at stage $k$ of the causal order to obtain measurement result $a_{l_k}$. Both of the above probabilities are conditional on a history, $H_{k-1}$, which contains all of the information about previous inputs, outputs and party order. In particular, the history $H_{k}=(h_1,...,h_k)$ is the ordered collection of triples $h_k=(l_k,a_{l_k},x_{l_k})$. The summations are performed over all possible next parties, excluding parties who have already acted, which are stored in the unordered sets $ \mathcal{L}_k=\{l_1,...,l_k\}$. To emphasise the symmetry between the terms we include $H_0$ and  $\mathcal{L}_0$, which are defined as empty sets, as no parties have acted at that point.
\end{definition}

\begin{definition}
The state of the system with a History $H_{k-1}$, at a time given by $t$, with the control set to trigger the action of party $l_k$ is given by 
\begin{align}
    \ket{\psi_{(l_k,t,H_{k-1})}}= (\ket{l_k}\bra{l_k}_c \otimes  \pi^{H_{k-1}}_{rf} \otimes I)U_t V U_{t-1}...V U_1 \ket{0}.
\end{align}
The projector onto the result and flag spaces is given by
$
     \pi^{H_{k-1}}_{rf} = \bigotimes_{i=1}^{N}\left(\pi^{H_{k-1}}_{r_{i}f_{i}}  \right),
$
where
\begin{align}
    \pi^{H_{k-1}}_{r_{i}f_{i}} = \begin{cases} 
    \ket{a_{i}}\bra{a_{i}}_{r_{i}} \otimes  \ket{1}\bra{1}_{f_{i}} & \text{ if } (i, a_i, x_i) \in H_{k-1},\\
   I_{r_{i}} \otimes \ket{0}\bra{0}_{f_{i}} & \text{ otherwise }.
    \end{cases} 
\end{align}
We also define the same state evolved to the end of protocol to be 
\begin{align}
     {\ket{\bar{\psi}_{(l_k,t,H_{k-1})}}}=VU_TVU_{T-1}...U_{t+1}V\ket{\psi_{((l_k,t,H_{k-1}))}}.
\end{align}
It will also be convenient for the proof to define $\ket{\psi_{(0,t,H_{k-1})}}$ and ${\ket{\bar{\psi}_{(0,t,H_{k-1})}}}$, which are the same as the above states, but with $l_k=0$ (i.e. the control in the `do nothing' setting).
\end{definition}

\begin{definition}
The state of the system with a History $H_{k}$, at a time given by $t$, in which party $l_k$ has just acted  is given by
\begin{align}
    \ket{\phi_{(l_k,t,H_{k})}}=(\ket{a_{l_k}}\bra{a_{l_k}}_{r_k} \otimes I)V\ket{\psi_{(l_k,t,H_{k-1})}}.
\end{align}
We also define the same state evolved to the end of protocol to be 
\begin{align}
    {\ket{\bar{\phi}_{(l_k,t,H_{k})}}}=VU_TVU_{T-1}...U_{t+1}\ket{\phi_{((l_k,t,H_{k}))}}.
\end{align}
It will also be convenient for the proof to define $\ket{\phi_{(0,t,H_k)}}=V\ket{\psi_{(0,t,H_k)}}$ and  ${\ket{\bar{\phi}_{(0,t,H_{k})}}}=VU_TVU_{T-1}...U_{t+1}\ket{\phi_{((0,t,H_{k}))}}.$
\end{definition}

\begin{definition} \label{def4} 
The probability for party $l_k$ to act next, given a history $H_{k-1}$ is given by:
\begin{align} \label{eq:probnext}
    p_k(l_k|H_{k-1})=\frac{\sum_{t_k=1}^{T}|\ket{\psi_{(l_k,t_k,H_{k-1})}}|^2}{\sum_{l_k' \notin \mathcal{L}_{k-1}}\sum_{t_k'=1}^{T}|\ket{\psi_{(l_k',t_k',H_{k-1})}}|^2}
\end{align}
\end{definition}

\begin{definition} \label{def5}
The probability for party $l_k$ to obtain the measurement result $a_{l_k}$, given a history $H_{k-1}$, and an input variable $x_{l_k}$ is given by:
\begin{align}\label{eq:probresult}
    p_k(a_{l_k}|H_{k-1},x_{l_k})=\frac{\sum_{t_k=1}^{T}|\ket{\phi_{(l_k,t_k,H_k)}}|^2}{\sum_{a'_{l_k} \in \mathcal{A}_{l_k}} \sum_{t_k'=1}^{T}|\ket{\phi_{(l_k,t_k',H_{k}')}}|^2}
\end{align}
where $H'_{k}=(H_{k-1}, (l_k, a'_{l_k}, x_{l_k}))$ (i.e. $H_k$ with $a_{l_k}$ replaced by $a'_{l_k}$).
\end{definition}

\begin{definition} The quantum protocol consists of preparing an initial state $\ket{0}$, then acting with an alternating sequence of unitaries $U_t$ that act on the system and the control, and unitaries $V$ that act on the system, results and flag spaces as specified by the control. The total unitary for the protocol is given by
\begin{align}
    \mathcal{U}=VU_{T}VU_{T-1}V...VU_{N}V...VU_{1}
\end{align}
where we note that for an $N$ party protocol, $T \geq N$. Finally, the results registers are measured in the computational basis, giving the outcome probability distribution
\begin{align} 
    p^{\mathrm{quantum}}(\vec{a}|\vec{x})=|(\ket{\vec{a}}\bra{\vec{a}}_r \otimes I)\mathcal{U}\ket{0}|^2. 
\end{align}
\end{definition}

With these definitions in place, we first  prove some useful orthogonality lemmas concerning the barred states.

\begin{lemma} \label{thm:lemma1} 
We have that 
\begin{align}
   \Braket{\bar{\psi}_{l',t',H} |  \bar{\psi}_{l,t,H} }=0 
\end{align}
unless $l=l'$ and $t'=t$.
\end{lemma}

Proof: consider first that $t=t'$ and $l\neq l'$. Then we have that $ \braket{\bar{\psi}_{l',t,H} |\bar{\psi}_{l,t,H}}=\braket{{\psi}_{l',t,H} |{\psi}_{l,t,H}}=0$, since $\ket{{\psi}_{l',t,H}}$ and $\ket{{\psi}_{l,t,H}}$ are orthogonal on the control $\mathcal{H}_c$. Next, consider that $t<t'$. Then $\braket{\bar{\psi}_{l',t',H} | \bar{\psi}_{l,t,H}}=\bra{{\psi}_{l',t',H}}U_{t'} V...U_{t+1}V\ket{{\psi}_{l,t,H}}=0$ since $V\ket{{\psi}_{l,t,H}}$ contains a raised $l$ flag that is not raised in $\bra{{\psi}_{l',t',H}}$, and there is no operator connecting the two which can lower this flag. The case with $t>t'$ follows from the $t<t'$ case by noting that $\Braket{\bar{\psi}_{l',t',H} |  \bar{\psi}_{l,t,H}}=\Braket{\bar{\psi}_{l,t,H} |  \bar{\psi}_{l',t',H}}^*$.

\begin{lemma} \label{thm:lemma2}
We have that 
\begin{align}
 \braket{\bar{\phi}_{l',t',H} | \bar{\phi}_{l,t,H}}=0
\end{align}
unless $l=l'$ and $t=t'$. 
\end{lemma}

Proof: consider first that $t=t'$ and $l \neq l'$. 
Then we have that $ \braket{\bar{\phi}_{l',t,H} |\bar{\phi}_{l,t,H}}=\braket{{\phi}_{l',t,H} |{\phi}_{l,t,H}}=0$, since $\ket{{\phi}_{l',t,H}}$ and $\ket{{\phi}_{l,t,H}}$ are orthogonal on the control $\mathcal{H}_c$. Next, consider that $t<t'$.
 Then $\braket{\bar{\phi}_{l',t',H} |\bar{\phi}_{l,t,H}}= \bra{{\phi}_{l',t',H}}V U_{t'} V ... U_{t+1}\ket{{\phi}_{l,t,H}}=0$, since the leftmost $V$ either raises a flag not in the history $H$, or the control at this point is set to zero, either of which will give the desired orthogonality. The case with $t>t'$ follows from the $t<t'$ case by noting that $\Braket{\bar{\phi}_{l',t',H} |  \bar{\phi}_{l,t,H}}=\Braket{\bar{\phi}_{l,t,H} |  \bar{\phi}_{l',t',H}}^*$.

\bigskip

We now move onto proving the main result. This will consist of four stages, the first concerns a cancellation within terms of the same causal order stage, which allows us to rewrite the causal model in a nice way. The second and third results concern the initial and final terms in the inductive proof. The former corresponds to the fact that `somebody has to measure first' in the quantum protocol, and the latter that the final term in the causal model has sufficient expressive power to capture the quantum measurement probabilities in their entirety. Finally, the fourth result concerns cancellations between terms at subsequent stages of the causal order. This leads to our main result which ties all of this together for a full proof that $p(\vec{a}|\vec{x})=|\bra{0}\mathcal{U}\ket{0}|^2$ is causal.  

\begin{result}\label{res::easycancel}
There is an equality between the numerator of the `who is next' type probabilities $p_k(l_k|H_{k-1})$, and the denominator of the `results' type probabilities $p_k(a_{l_k}|H_{k-1},x_{l_k})$, allowing us to  write the product of these probabilities in a nice way as
\begin{align}
    p_k(l_k|H_{k-1}) p_k(a_{l_k}|x_{l_k},H_{k-1}) = \frac{\sum_{t_k=1}^{T}|\ket{\phi_{(l_k,t_k,H_{k})}}|^2}{\sum_{l'_k \notin \mathcal{L}_{k-1}}\sum_{t'_k=1}^{T}|\ket{\psi_{(l'_k,t_k',H_{k-1})}}|^2}.
\end{align}
\end{result}. 

Proof: Starting with the denominator of the `results' probability
\begin{align}
    \sum_{a'_{l_k} \in \mathcal{A}_{l_k}} \sum_{t_k=1}^{T}\left|\ket{\phi_{(l_k,t_k,H'_{k})}}\right|^2&=  \sum_{a'_{l_k} \in \mathcal{A}_{l_k}}\sum_{t_k=1}^{T}\left|(\ket{a'_{l_k}}\bra{a'_{l_k}}_{r_k} \otimes  \mathcal{I})V\ket{\psi_{(l_k,t_k,H_{k-1})}}\right|^2 \nonumber \\
    &=\sum_{t_k=1}^{T}\left|\sum_{a'_{l_k} \in \mathcal{A}_{l_k}} (\ket{a'_{l_k}}\bra{a'_{l_k}}_{r_k} \otimes \mathcal{I})V\ket{\psi_{(l_k,t_k,H_{k-1})}}\right|^2. \nonumber \\
        &=\sum_{t_k=1}^{T}|V\ket{\psi_{(l_k,t_k,H_{k-1})}}|^2\\
    &=\sum_{t_k=1}^{T}|\ket{\psi_{(l_k,t_k,H_{k-1})}}|^2,
\end{align}
we obtain the numerator of the `who is next' probabilities. In the second line we have used orthogonality on the results register, in the third line we have used the fact that after a measurement by party $l_k$, some result in $\mathcal{A}_{l_k}$ must have been obtained, and in the final line we have used unitarity. Using this to cancel the numerator of \eqref{eq:probnext} with the denominator of \eqref{eq:probresult} we obtain the desired result.  

\begin{result}\label{res::firstterm}
The denominator of the first term $p_1(l_1 | H_0)$ satisfies 
\begin{align}
    \sum_{l_1 \notin \mathcal{L}_0} \sum_{t_1=1}^{T}|\ket{\psi_{(l_1,t_1,H_0)}}|^2=1.
\end{align}
\end{result}
Proof: by first using unitarity and then Lemma \ref{thm:lemma1} we have 
\begin{align}\label{eq::res2eq1}
    \sum_{l_1 \notin \mathcal{L}_0} \sum_{t_1=1}^{T}|\ket{\psi_{(l_1,t_1,H_0)}}|^2 
    =\sum_{l_1 \notin \mathcal{L}_0}\sum_{t_1=1}^{T}|\ket{\bar{\psi}_{(l_1,t_1,H_0})}|^2 \nonumber \\
    =|\sum_{l_1 \notin \mathcal{L}_0}\sum_{t_1=1}^{T}\ket{\bar{\psi}_{(l_1,t_1,H_0})}|^2.
\end{align}
To simplify this further, consider evolving the state  $\ket{\psi_{0,t_1-1,H_0}}$ forward for a full time-step using $U_{t_1}V$. As the control is in state $0$, no measurement occurs during $V$, and the unitary $U_1$ creates a superposition in which the control takes any possible state. Symbolically, 
\begin{align}
    U_{t_1}V \ket{\psi_{0,t_1-1,H_0}}=\sum_{l_1 \notin \mathcal{L}_0}\ket{\psi_{(l_1,t_1,H_0})}+\ket{\psi_{(0,t_1,H_0})}.
\end{align}
By applying $VU_TV \ldots U_{t_1 + 1} V$ to this equation, we can obtain a similar form for the barred states,
\begin{align}
    \ket{\bar{\psi}_{0,t_1-1,H_0}}= \sum_{l_1 \notin \mathcal{L}_0}\ket{\bar{\psi}_{(l_1,t_1,H_0})}+\ket{\bar{\psi}_{(0,t_1,H_0})}.
\end{align}
We can rearrange this equation and substitute for the sum over $l_1$ on the right-hand side of  \eqref{eq::res2eq1} to obtain
\begin{align}
     \sum_{l_1 \notin \mathcal{L}_0} \sum_{t_1=1}^{T}|\ket{\psi_{(l_1,t_1,H_0)}}|^2 = |\sum_{t_1=1}^{T}\left( \ket{\bar{\psi}_{(0,t_1-1,H_0})}-\ket{\bar{\psi}_{(0,t_1,H_0)}}\right)|^2
\end{align}
By expanding the summation on the right-hand side we find that only the first and last terms remain, giving 
\begin{align}
     \sum_{l_1 \notin \mathcal{L}_0} \sum_{t_1=1}^{T}|\ket{\psi_{(l_1,t_1,H_0)}}|^2    =|\ket{\bar{\psi}_{(0,0,H_0})}-\ket{\bar{\psi}_{(0,T,H_0})}|^2
\end{align}
Note that it is impossible by the requirements of our protocol that no-one has measured by time $t=T$. Such a scenario would violate the assumption that there are exactly $N$ flags raised at the end of the protocol. Therefore, $\ket{\bar{\psi}_{(0,T,H_0})}=0$, and we find
\begin{align}
       \sum_{l_1 \notin \mathcal{L}_0} \sum_{t_1=1}^{T}|\ket{\psi_{(l_1,t_1,H_0)}}|^2 =  |\ket{\bar{\psi}_{(0,0,H_0})}|^2=|\mathcal{U} \ket{0}|^2=1
\end{align}
as desired.

\begin{result}\label{res::lastterm}
The outcome statistics in the numerator of the final term in the causal probabilistic model represent the quantum probabilities arising from the protocol. In other words,
\begin{align}
    \sum_{l_N \in \mathcal{L}_N}\sum_{t_N=1}^T |\ket{\phi_{(l_N,t_N,H_N)}}|^2= |\left(\ket{\vec{a}}\bra{\vec{a}} \otimes I \right) \mathcal{U}\ket{0}|^2
\end{align}
\end{result}
Proof:
Firstly, note that by  unitarity and Lemma \ref{thm:lemma2}  we have that
\begin{align}\label{eq::res3eq1}
    \sum_{l_N \in \mathcal{L}_N}\sum_{t_N=1}^T |\ket{\phi_{(l_N,t_N,H_N)}}|^2 &=   \sum_{l_N \in \mathcal{L}_N}\sum_{t_N=1}^T |\ket{\bar{\phi}_{(l_N,t_N,H_N)}}|^2\nonumber \\
   &= |\sum_{l_N \in \mathcal{L}_N}\sum_{t_N=1}^T \ket{\bar{\phi}_{(l_N,t_N,H_N)}}|^2.
\end{align}
The history $H_N$ represents a case in which all parties have already measured. At time $t<T$, what are the possible ways that this history can be filled? Either, nobody has measured in the previous time-step, or the last party to be filled into the history (subject to the requirement every party must enter the history exactly once) has just measured. In any case, evolving the linear combination of these states forward a time-step must produce a state at $t+1$ which contains an empty control (i.e., no-one else is left to measure, so don't trigger them!). This gives us the key relation 
\begin{align}
    VU_{t+1} \left( \sum_{l_N \in \mathcal{L}_N} \ket{\phi_{(l_N,t,H_N})}+\ket{\phi_{(0,t,H_N)}} \right)=\ket{\phi_{(0,t+1,H_N})}
\end{align}
which holds for $t < T$. Applying $VU_TV \ldots U_{t+2} $ we obtain 
\begin{align}
  \left( \sum_{l_N \in \mathcal{L}_N} \ket{\bar{\phi}_{(l_N,t,H_N})}+\ket{\bar{\phi}_{(0,t,H_N)}} \right)=\ket{\bar{\phi}_{(0,t+1,H_N})}
\end{align}
which we can then rearrange to get
\begin{align}
   \sum_{l_N \in \mathcal{L}_N} \ket{\bar{\phi}_{(l_N,t,H_N)}}=\ket{\bar{\phi}_{(0,t+1,H_N)}}-\ket{\bar{\phi}_{(0,t,H_N)}}.
\end{align}
By separating out the $t_N=T$ term in equation \eqref{eq::res3eq1} and then substituting this in the remaining terms , we find that
\begin{align}
   \sum_{l_N \in \mathcal{L}_N}\sum_{t_N=1}^T |\ket{\phi_{(l_N,t_N,H_N)}}|^2 &=    |\sum_{l_N \in \mathcal{L_N}}\ket{\bar{\phi}_{(l_N,T,H_N)}}+\sum_{t_N=1}^{T-1}\sum_{l_N \in \mathcal{L_N}}\ket{\bar{\phi}_{(l_N,t_N,H_N)}}|^2 \nonumber \\
  & = |\sum_{l_N \in \mathcal{L_N}}\ket{\bar{\phi}_{(l_N,T,H_N)}}+\sum_{t_N=1}^{T-1}\left(\ket{\bar{\phi}_{(0,t_N+1,H_N)}}-\ket{\bar{\phi}_{(0,t_N,H_N)}} \right) |^2 \nonumber \\
  &=  |\sum_{l_N \in \mathcal{L_N}}\ket{\bar{\phi}_{(l_N,T,H_N)}}+\ket{\bar{\phi}_{(0,T,H_N)}}-\ket{\bar{\phi}_{(0,1,H_N)}}|^2.
\end{align}
Now we note that $\ket{\bar{\phi}_{(0,1,H_N)}}=0$ since  it would be impossible for all parties to have measured in one time-step, and for the control to be in the zero state. Then we note that $\sum_{l_N \in \mathcal{L_N}}\ket{\bar{\phi}_{(l_N,T,H_N)}}+\ket{\bar{\phi}_{(0,T,H_N)}}= (\pi^{H_{N}}_{rf} \otimes I )\mathcal{U}\ket{0}$, which is to say that these are simply the possible states at the end of the protocol, containing the measurement results we want to calculate the probabilities for in the history.  Therefore
\begin{align}
 \sum_{l_N \in \mathcal{L}_N}\sum_{t_N=1}^T |\ket{\phi_{(l_N,t_N,H_N)}}|^2 &=     |(\pi^{H_{N}}_{rf} \otimes I )\mathcal{U} \ket{0}|^2 \nonumber \\
 &=
    |( \ket{\vec{a}}\bra{\vec{a}} \otimes I) \mathcal{U}\ket{0}|^2
\end{align}
as desired. 

\begin{result}\label{res::diagcancel}
This is a technical result which establishes an equality between the states after measurement at causal order stage $k$ and the states before measurement at the next stage of the causal order. Namely, for $1 \leq k<N$ that:
\begin{align}
    \sum_{l_k \in \mathcal{L}_k}\sum_{t=1}^{T}|\ket{\phi_{(l_k, t, H_k)}}|^2=\sum_{t'=1}^{T}\sum_{l'_{k+1} \notin \mathcal{L}_k} |\ket{{\psi}_{(l'_{k+1},t',H_{k})}}|^2.
\end{align}
\end{result}
Proof: Firstly, by  unitarity and Lemma \ref{thm:lemma2} we have that
\begin{align}\label{eq::res4eq1}
\sum_{l_k \in \mathcal{L}_k}\sum_{t=1}^{T}|\ket{\phi_{(l_k, t, H_k)}}|^2 &=\sum_{l_k \in\mathcal{L}_k}\sum_{t=1}^{T}|\ket{\bar{\phi}_{(l_k, t, H_k)}}|^2 \nonumber\\
&=|\sum_{l_k \in \mathcal{L}_k}\sum_{t=1}^{T}\ket{\bar{\phi}_{(l_k, t, H_k)}}|^2.
\end{align}
Consider time-evolving a state just after the $t^{\textrm{th}}$ measurement step, in which history $H_k$ has been obtained (either by the last party just having measured, or by all parties in $H_k$ having measured previously), by $U_{t+1}$. This links states of the form $\ket{\phi_{(l_k,t,H_k)}}$ and $\ket{\psi_{(l'_k,t+1,H_k)}}$ via 
\begin{align}
    U_{t+1} \left( \sum_{l_k \in \mathcal{L}_k} \ket{\phi_{(l_k,t,H_k)}}+\ket{\phi_{(0,t,H_k)}} \right)=\sum_{l'_{k+1}
    \notin \mathcal{L}_k}\ket{\psi_{(l'_{k+1},t+1,H_{k})}} + \ket{\psi_{(0,t+1,H_{k})}}
\end{align}
for $t<T$. Applying $VU_T V \ldots U_{t+2} V$ we obtain a very similar result for barred states;
\begin{align}
    \sum_{l_k \in \mathcal{L}_k} \ket{\bar{\phi}_{(l_k,t,H_k)}} +\ket{\bar{\phi}_{(0,t,H_{k})}} =\sum_{l'_{k+1} \notin \mathcal{L}_k}\ket{\bar{\psi}_{(l'_{k+1},t+1,H_{k})}} + \ket{\bar{\psi}_{(0,t+1,H_{k})}}. 
\end{align}
We also note that $\ket{\phi_{(0,t,H_{k})}}=V\ket{\psi_{(0,t,H_{k})}}$ and hence that $\ket{\bar{\phi}_{(0,t,H_{k})}}=\ket{\bar{\psi}_{(0,t,H_{k})}}$. Making this substitution and rearranging a little we get 
\begin{align}\label{eq::res4eq2}
    \sum_{l_k \in \mathcal{L}_k} \ket{\bar{\phi}_{(l_k,t,H_k)}}  =\sum_{l'_{k+1} \notin \mathcal{L}_k}\ket{\bar{\psi}_{(l'_{k+1},t+1,H_{k})}} + \ket{\bar{\psi}_{(0,t+1,H_{k})}} - \ket{\bar{\psi}_{(0, t, H_k)}}. 
\end{align}
By  substituting \eqref{eq::res4eq2} into \eqref{eq::res4eq1} for $t<T$, we arrive at
\begin{align}
  \sum_{l_k \in \mathcal{L}_k}\sum_{t=1}^{T}\left|\ket{\phi_{(l_k, t, H_k)}}\right|^2  &=\left|\sum_{l_k \in \mathcal{L}_k}\ket{\bar{\phi}_{(l_k, T, H_k)}} + \sum_{t=1}^{T-1} \left( \sum_{l'_{k+1} \notin \mathcal{L}_k}\ket{\bar{\psi}_{(l'_{k+1},t+1,H_{k})}} + \ket{\bar{\psi}_{(0,t+1,H_{k})}}-\ket{\bar{\psi}_{(0,t,H_{k})}} \right) \right|^2 \nonumber \\
    &=\left|\sum_{l_k \in \mathcal{L}_k}\ket{\bar{\phi}_{(l_k, T, H_k)}} + \sum_{t=1}^{T-1}  \sum_{l'_{k+1} \notin \mathcal{L}_k}\ket{\bar{\psi}_{(l'_{k+1},t+1,H_{k})}} + \ket{\bar{\psi}_{(0,T,H_{k})}}-\ket{\bar{\psi}_{(0,1,H_{k})}}  \right|^2 \label{eq:result4eqn}
\end{align}
where for the sums over time in the last two terms only the states with maximal and minimal times remain. 

Given that $k<N$ and all parties must have measured by the end of the protocol, it must be the case that $\ket{\bar{\phi}_{(l_k, T, H_k)}}=0$ and $\ket{\bar{\psi}_{(0,T,H_{k})}}=0$. Also as $k \geq 1$ it must be the case that   $\ket{\bar{\psi}_{(0,1,H_{k})}}=0$ and 
$\ket{\bar{\psi}_{(l'_{k+1},1, H_{k})}}=0$, as these states are just before the first measurement and hence must have no history. Using these results in equation \eqref{eq:result4eqn} and setting $t'=t+1$, we obtain 
\begin{align}
    \sum_{l_k \in \mathcal{L}_k}\sum_{t=1}^{T}|\ket{\phi_{(l_k, t, H_k)}}|^2=\left| \sum_{t'=1}^{T}\sum_{l'_{k+1} \notin \mathcal{L}_k} \ket{\bar{\psi}_{(l'_{k+1},t',H_{k})}} \right|^2.
\end{align}
Finally, using Lemma \ref{thm:lemma1} and unitarity we arrive at 
\begin{align}
    \sum_{l_k \in \mathcal{L}_k}\sum_{t=1}^{T}|\ket{\phi_{(l_k, t, H_k)}}|^2&= \sum_{t'=1}^{T}\sum_{l'_{k+1} \notin \mathcal{L}_k} \left| \ket{\bar{\psi}_{(l'_{k+1},t',H_{k})}} \right|^2 \nonumber \\
    &= \sum_{t'=1}^{T}\sum_{l'_{k+1} \notin \mathcal{L}_k} \left| \ket{{\psi}_{(l'_{k+1},t',H_{k})}} \right|^2 
\end{align}
as required.

\begin{result}
We will now show that the results of the quantum protocol can be replicated by a causal process. In particular
\begin{align}
         p(\vec{a}|\vec{x})=|\left( \ket{\vec{a}}\bra{\vec{a}} \otimes I \right)\mathcal{U}\ket{0}|^2&=\nonumber\\
         \sum_{l_1 \notin \mathcal{L}_0} \sum_{l_2 \notin \mathcal{L}_1}& ... \sum_{l_N \notin \mathcal{L}_N-1} p_1(l_1|H_0)p_1(a_{l_1}|H_0,x_{l_1}) p_2(l_2|H_1)p_2(a_{l_1}|H_1,x_{l_1})...
    p_N(l_N|H_{N-1})p_N(a_{l_N}|H_{N-1},x_{l_N})
\end{align}
and as such, the outcome statistics $p(\vec{a}|\vec{x})$ cannot violate a causal inequality.
\end{result}

Proof: Firstly, substituting definitions \ref{def4} and \ref{def5} into the causal model \eqref{def::modelsi}, and then using Result \ref{res::easycancel}, we can re-write the probability distribution for the entire causal model as 
\begin{align}\label{eq::modelredefined}
     p(\vec{a}|\vec{x})=\sum_{l_1 \notin \mathcal{L}_0} \sum_{l_2 \notin \mathcal{L}_1} ... \sum_{l_N \notin \mathcal{L}_N-1} \frac{\sum_{t_1=1}^{T}|\ket{\phi_{(l_1,t_1,H_{1})}}|^2}{\sum_{l_1' \notin \mathcal{L}_{0}}\sum_{t_1'=1}^{T}|\ket{\psi_{(l_1',t_1',H_{0})}}|^2} &\frac{\sum_{t_2=1}^{T}|\ket{\phi_{(l_2,t_2,H_{2})}}|^2}{\sum_{l_2' \notin \mathcal{L}_{1}}\sum_{t_2'=1}^{T}|\ket{\psi_{(l_2',t_2',H_{1})}}|^2}...\nonumber \\
     ...&\frac{\sum_{t_N=1}^{T}|\ket{\phi_{(l_N,t_N,H_N)}}|^2}{\sum_{l_N' \notin \mathcal{L}_{N-1}}\sum_{t_N'=1}^{T}|\ket{\psi_{(l_N',t_N',H_{N-1})}}|^2}
\end{align}

Let us begin by performing a simple reshuffling of \eqref{eq::modelredefined}'s numerators and denominators, by writing the denominator of the term associated to causal order stage $k$ as the denominator of the term associated to $k-1$.
\begin{align}
     p(\vec{a}|\vec{x})=\frac{1}{\sum_{l_1' \notin \mathcal{L}_{0}}\sum_{t_1'=1}^{T}|\ket{\psi_{(l_1',t_1',H_{0})}}|^2}\sum_{l_1 \notin \mathcal{L}_0} \sum_{l_2 \notin \mathcal{L}_1} ... \sum_{l_N \notin \mathcal{L}_N-1}&\frac{\sum_{t_1=1}^{T}|\ket{\phi_{(l_1,t_1,H_{1})}}|^2}{\sum_{l_2' \notin \mathcal{L}_{1}}\sum_{t_2'=1}^{T}|\ket{\psi_{(l_2',t_2',H_{1})}}|^2}
     \frac{\sum_{t_2=1}^{T}|\ket{\phi_{(l_2,t_2,H_{2})}}|^2}{\sum_{l_3' \notin \mathcal{L}_{2}}\sum_{t_3'=1}^{T}|\ket{\psi_{(l_3',t_3',H_{2})}}|^2}...\nonumber\\ 
   &...\frac{\sum_{t_{k-1}=1}^{T}|\ket{\phi_{(l_{k-1},t_{k-1},H_{k-1})}}|^2}{\sum_{l_k' \notin \mathcal{L}_{k-1}}\sum_{t_k'=1}^{T}|\ket{\psi_{(l_k',t_k',H_{k-1})}}|^2}...
     \sum_{t_N=1}^{T}|\ket{\phi_{(l_N,t_N,H_N)}}|^2
\end{align}
by using Result \ref{res::firstterm} we have 
\begin{align}
     p(\vec{a}|\vec{x})&=\sum_{l_1 \notin \mathcal{L}_0} \sum_{l_2 \notin \mathcal{L}_1} ... \sum_{l_N \notin \mathcal{L}_N-1}\frac{\sum_{t_1=1}^{T}|\ket{\phi_{(l_1,t_1,H_{1})}}|^2}{\sum_{l_2' \notin \mathcal{L}_{1}}\sum_{t_2'=1}^{T}|\ket{\psi_{(l_2',t_2',H_{1})}}|^2}\frac{\sum_{t_2=1}^{T}|\ket{\phi_{(l_2,t_2,H_{2})}}|^2}{\sum_{l_3' \notin \mathcal{L}_{2}}\sum_{t_3'=1}^{T}|\ket{\psi_{(l_3',t_3',H_{2})}}|^2} 
   ...
     \sum_{t_N=1}^{T}|\ket{\phi_{(l_N,t_N,H_N)}}|^2
\end{align}

now note that we can rewrite the leftmost sum as 
\begin{align} 
\sum_{l_1 \notin \mathcal{L}_0} = \sum_{\mathcal{L}_1} \sum_{l_1 \in \mathcal{L}_1}, 
\end{align} 
where the sum over $\mathcal{L}_1$ is over all singleton sets $\{l_1\}$ (and hence has $N$ terms), and the subsequent sum over ${l_1 \in \mathcal{L}_1}$ contains just a single term. 

We can use this to rewrite the probability distribution as 
\begin{align}
  p(\vec{a}|\vec{x})&=\sum_{\mathcal{L}_1}\frac{\sum_{l_1 \in \mathcal{L}_1}\sum_{t_1=1}^{T}|\ket{\phi_{(l_1,t_1,H_{1})}}|^2}{\sum_{l_2' \notin \mathcal{L}_{1}}\sum_{t_2'=1}^{T}|\ket{\psi_{(l_2',t_2',H_{1})}}|^2} \sum_{l_2 \notin \mathcal{L}_1} ... \sum_{l_N \notin \mathcal{L}_N-1}\frac{\sum_{t_2=1}^{T}|\ket{\phi_{(l_2,t_2,H_{2})}}|^2}{\sum_{l_3' \notin \mathcal{L}_{2}}\sum_{t_3'=1}^{T}|\ket{\psi_{(l_3',t_3',H_{2})}}|^2} 
   ...
     \sum_{t_N=1}^{T}|\ket{\phi_{(l_N,t_N,H_N)}}|^2
\end{align}
where we have used the fact that the first numerator and denominator do not depend on $\{l_2, \ldots l_N\}$. By  application of Result \ref{res::diagcancel} this is just 
\begin{align}
  p(\vec{a}|\vec{x})&=\sum_{\mathcal{L}_1} \sum_{l_2 \notin \mathcal{L}_1} ... \sum_{l_N \notin \mathcal{L}_N-1}\frac{\sum_{t_2=1}^{T}|\ket{\phi_{(l_2,t_2,H_{2})}}|^2}{\sum_{l_3' \notin \mathcal{L}_{2}}\sum_{t_3'=1}^{T}|\ket{\psi_{(l_3',t_3',H_{2})}}|^2} 
   ...
     \sum_{t_N=1}^{T}|\ket{\phi_{(l_N,t_N,H_N)}}|^2
\end{align}
we can iterate the same process again using 
\begin{align} 
\sum_{\mathcal{L}_1} \sum_{l_2 \notin \mathcal{L}_1} = \sum_{\mathcal{L}_2} \sum_{l_2 \in \mathcal{L}_2}.
\end{align}
The left-hand side corresponds to first picking $l_1$ (with $N$ possibilities) and then picking a different $l_2$ ($N-1$ possibilities), whereas the right-hand side corresponds to first picking a pair of distinct labs $\mathcal{L}_2$ (with $N(N-1)/2$ possibilities) and then picking which of them was last ($2$ possibilities). This gives 
\begin{align}
      p(\vec{a}|\vec{x})&=\sum_{\mathcal{L}_2} \sum_{l_2 \in \mathcal{L}_2} ... \sum_{l_N \notin \mathcal{L}_N-1}\frac{\sum_{t_2=1}^{T}|\ket{\phi_{(l_2,t_2,H_{2})}}|^2}{\sum_{l_3' \notin \mathcal{L}_{2}}\sum_{t_3'=1}^{T}|\ket{\psi_{(l_3',t_3',H_{2})}}|^2} 
   ...
     \sum_{t_N=1}^{T}|\ket{\phi_{(l_N,t_N,H_N)}}|^2
\end{align}
which is just 
\begin{align}
    p(\vec{a}|\vec{x})&=\sum_{\mathcal{L}_2} \frac{\sum_{l_2 \in \mathcal{L}_2}\sum_{t_2=1}^{T}|\ket{\phi_{(l_2,t_2,H_{2})}}|^2}{\sum_{l_3' \notin \mathcal{L}_{2}}\sum_{t_3'=1}^{T}|\ket{\psi_{(l_3',t_3',H_{2})}}|^2} 
    \sum_{l_3 \notin \mathcal{L}_2}
    ... \sum_{l_N \notin \mathcal{L}_N-1} \frac{\sum_{t_3=1}^T |\ket{\phi_{l_3,t_3,H_3}}|^2}{\sum_{l_4' \notin \mathcal{L}_{3}}\sum_{t_4'=1}^{T}|\ket{\psi_{(l_4',t_4',H_{3})}}|^2}
   ...
     \sum_{t_N=1}^{T}|\ket{\phi_{(l_N,t_N,H_N)}}|^2
\end{align}
application of Result \ref{res::diagcancel} leads to another cancellation, so that we may write now 
\begin{align}
    p(\vec{a}|\vec{x})&=\sum_{\mathcal{L}_2}
    \sum_{l_3 \notin \mathcal{L}_2}... \sum_{l_N \notin \mathcal{L}_N-1}\frac{\sum_{t_3=1}^T |\ket{\phi_{l_3,t_3,H_3}}|^2}{\sum_{l_4' \notin \mathcal{L}_{3}}\sum_{t_4'=1}^{T}|\ket{\psi_{(l_4',t_4',H_{3})}}|^2}
   ...
     \sum_{t_N=1}^{T}|\ket{\phi_{(l_N,t_N,H_N)}}|^2
\end{align}
We can then iterate this process by applying the general result that  
\begin{align} 
\sum_{\mathcal{L}_k} \sum_{l_{k+1} \notin \mathcal{L}_k} = \sum_{\mathcal{L}_{k+1}} \; \sum_{l_{k+1} \in \mathcal{L}_{k+1}},
\end{align}
and cancelling one of the numerators and denominators using Result \ref{res::diagcancel} until we are left with 
 the final term, 
\begin{align}
     p(\vec{a}|\vec{x})&=\sum_{\mathcal{L}_N}\sum_{l_n \in \mathcal{L}_N}\sum_{t_N=1}^{T}|\ket{\phi_{(l_N,t_N,H_N)}}|^2
\end{align}
the summation $\sum_{\mathcal{L}_N}=1$, as the only term corresponds to $\mathcal{L}_N = \{1,2, \ldots N\}$. Application of Result \ref{res::lastterm} then shows that this causal model indeed reproduces the quantum probabilities.i.e. that 
\begin{align}
    p(\vec{a}|\vec{x})=|\left( \ket{\vec{a}} \bra{\vec{a}}\otimes I \right) \mathcal{U}\ket{0}|^2
\end{align}
as desired.

\section{Example}
 We now give an example of how our results apply in practice, based on the quantum switch \cite{Chiribella2013}. This involves using a quantum control  to determine the order in which two operations are applied to another quantum system. The switch can be modelled in  number of different ways (e.g. as a process matrix that exhibits causal non-separability \cite{Branciard2016}) but here the basic idea is to prepare a superposition state of the form 
\begin{equation} 
\frac{1}{\sqrt{2}} \left( \ket{1}_c \otimes U_A U_B \ket{0}_{sfr} + \ket{2}_c \otimes U_B U_A \ket{0}_{sfr} \right) 
\end{equation} 
where $U_A$ and $U_B$ are unitary transformations  by Alice and Bob (representing their measurements). If a third party, Charlie, measures the control in a basis consisting of superpositions of $\ket{1}$ and $\ket{2}$ this will introduce interference between the two causal orders in which either Alice or Bob goes first. It has already been shown that this simple setup cannot be used to violate a causal inequality \cite{Arajo2015, Oreshkov2012}. However, it is instructive to see how it fits into our framework. 
\begin{figure}[h]\
\centering
\Qcircuit @C=1.5em @R=1.2em {
\\
&\ustick{\ket{0}_{c}}&\qw & \gate{U_1} & \ctrl{1} & \gate{U_2} & \ctrl{1}  & \multigate{2}{U_3} & \ctrl{1} & \qw \\
&\ustick{\ket{0}_{s_1}}&\qw&\qw & \multigate{3}{V} & \qw & \multigate{3}{V} & \ghost{U_3} & \multigate{3}{V}& \qw \\
&\ustick{\ket{0}_{s_{2}}}&\qw & \qw & \ghost{V} & \qw & \ghost{V} & \ghost{U_3} & \ghost{V}& \qw \\
&\ustick{\ket{0}_{r}}&\qw & \qw & \ghost{V} & \qw & \ghost{V} & \qw & \ghost{V}& \qw \\
&\ustick{\ket{0}_{f}}&\qw & \qw & \ghost{V} & \qw & \ghost{V} & \qw & \ghost{V}& \qw &\\
}\caption{Realisation of the quantum switch in our framework through a quantum circuit.}
\label{fig:newcirc}
\end{figure}
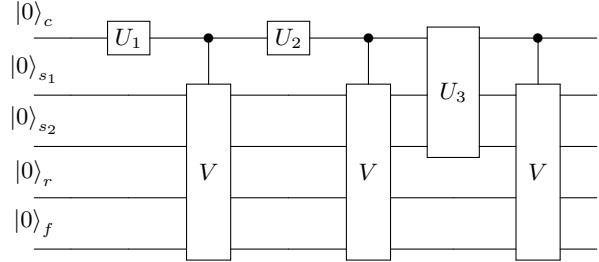

Because parties cannot directly measure the control in our framework, we transfer the state of the control into the system before Charlie's measurement, and split the system into two qubits to facilitate this. Overall, the circuit we consider is shown in figure  \ref{fig:newcirc}, where

\begin{align}
    U_1\ket{0}_c &=\frac{1}{\sqrt{2}} \left( \ket{1}_c + \ket{2}_c \right),\nonumber\\
    U_2\ket{1}_c &= \ket{2}_c \nonumber\\ 
    U_2 \ket{2}_c &= \ket{1}_c \nonumber\\
    U_3\ket{1}_c\ket{\psi}_{s_1}\ket{0}_{s_2} &= \ket{3}_c \ket{0}_{s_1}\ket{\psi}_{s_2}, \nonumber\\
    U_3\ket{2}_c\ket{\psi}_{s_1}\ket{0}_{s_2} &= \ket{3}_c \ket{1}_{s_1}\ket{\psi}_{s_2},
\end{align} 
and $V$ is as given in equation 4 of the main body text. 

By considering the outcome statistics generated by this switch setup, we find that they differ from those which would be obtained from an equal mixture of the causal orders  $A\rightarrow B \rightarrow C$ and $B\rightarrow A \rightarrow C$, due to the presence of interference. 

We proceed with an explicit calculation for the setup in figure \ref{fig:newcirc}. We assume that all parties have two possible measurements, hence their input variables $x,y,z$ are bits. When their input bit is zero, they 
 measure the first part of the system in the computational basis and output the result in $a,b,c$. When their input bit is one, they instead measure the first part of the system in the Fourier basis (composed of the  states $\ket{\pm} = \frac{1}{\sqrt{2}} \left( \ket{0} \pm \ket{1} \right)$, and output  zero if they obtain the state $\ket{+}$ and one if they obtain the state $\ket{-}$. All of these measurements are implemented via unitary operations between the system and result register. e.g. 
 \begin{align} 
 V_{s_1, r_1}(x_1=1) &= \ket{+}\bra{+}_{s_1} \otimes I_{r_1} + \ket{-}\bra{-}_{s_1} \otimes \left( \ket{0}\bra{1}_{r_1} + \ket{1}\bra{0}_{r_1}\right)
 \end{align}

Consider the case where the input variables are $x=0,y=1,z=1$. After some calculation, we find the state at the end of the protocol to be 
\begin{align}
\ket{3}_c\bigg( &\ket{+}_{s_1}(\frac{1}{2\sqrt{2}}\ket{+}_{s_2}+\frac{1}{4}\ket{0}_{s_2}) \ket{000}_r 
+\ket{-}_{s_1}(\frac{1}{4}\ket{0}_{s_2}- \frac{1}{2\sqrt{2}}\ket{+}_{s_2}) \ket{001}_r \nonumber\\
+&\ket{+}_{s_1}(\frac{1}{2\sqrt{2}}\ket{-}_{s_2}+\frac{1}{4}\ket{0}_{s_2}) \ket{010}_r 
+\ket{-}_{s_1}(\frac{1}{4}\ket{0}_{s_2}- \frac{1}{2\sqrt{2}}\ket{-}_{s_2}) \ket{011}_r \nonumber\\
+&\frac{1}{4}\ket{+}_{s_1}\ket{1}_{s_2} \ket{100}_r
+\frac{1}{4}\ket{-}_{s_1}\ket{1}_{s_2} \ket{101}_r
-\frac{1}{4}\ket{+}_{s_1}\ket{1}_{s_2} \ket{110}_r
-\frac{1}{4}\ket{-}_{s_1}\ket{1}_{s_2} \ket{111}_r\bigg)\ket{111}_f,
\end{align}
where we adopt the convention that $\ket{000}_r =\ket{a=0, b=0, c=0}_r $, et cetera. 
The probabilities to observe different outcomes in this measurement setting can then be obtained from this state. For example, $p(000|011)=|\frac{1}{2\sqrt{2}}\ket{+}_{s_1}\ket{+}_{s_2}+\frac{1}{4}\ket{+}_{s_1}\ket{0}_{s_2}|^2=5/16$. Such probabilities notably involve interference between different causal orders. In particular, they differ from those which would be obtained in the naive classical case, in which we first flip a coin to determine which of the two causal orders we will place ourselves in, and then perform the measurements in this causal order. We now calculate this `naive causal' probability $p^{\text{nc}}(000|011)$. One half of the time, when we are in the causal order $A\rightarrow B \rightarrow C$, Alice measures $0$ with certainty and Bob, and Charlie have each a $50:50$ chance to measure either $0$ or $1$. The other half of the time, we are in the causal order $B\rightarrow A \rightarrow C$ all parties have a $50:50$ chance to measure either $0$ or $1$ (since Bob's measurement in the Fourier basis, which occurs first, now makes Alice completely uncertain of her outcome). Putting this all together we find $p^{\text{nc}}(000|011)= \frac{1}{2} \times 1 \times \frac{1}{2} \times  \frac{1}{2} + \frac{1}{2} \times\frac{1}{2} \times \frac{1}{2} \times \frac{1}{2} = 3/16 \neq p(000|011)$.

Nevertheless, our results show that we \textit{can} find some classical causal model which generates the same outcome distribution $p(abc|xyz)$ as the quantum case. Let's do this explicitly, to see how our proof translates in practice.

We find by direct substitution into the definitions 4, and 5 in the main body that:
\begin{align}
   &  p_1(l_1=\text{Alice}|H_0)=1/2 \quad\nonumber\\ 
   &\qquad\qquad p_1(l_1=\text{Bob}|H_0)=1/2 \nonumber\\
    & p_2(l_2=\text{Bob}|H_1=(1,0,0))=1 \quad \nonumber \\
    &\qquad \qquad p_2(l_2=\text{Alice}|H_1=(2,0,1))=1\nonumber \\
    & p_3(l_3=\text{Charlie}|H_2=((1,0,0),(2,0,1))=1 \quad \nonumber \\ 
    & \qquad \qquad p_3(l_3=\text{Charlie}|H_2=((2,0,1),(1,0,0))=1\nonumber\\
    &p_1(a=0|H_0,x=0)=1 \quad \nonumber\\
    &\qquad\qquad p_1(b=0|H_0,y=1)=1/2 \nonumber \\
    & p_2(a=0|H_1=(1,0,0),x=0)=1/2 \quad \nonumber\\
    & \qquad \qquad p_2(b=0|H_1=(2,0,1),y=1)=1/2\nonumber\\
\end{align}
which are all the same as the naive classical case. However, the  results-type probability for Charlie differs from the naive case. In particular we find 
\begin{align} 
    & p_3(c=0|H_2=((1,0,0),(2,0,1), z=1)=\frac{5}{6} \quad \nonumber \\ 
    & \qquad \qquad p_3(c=0|H_2=((2,0,1),(1,0,0), z=1)=\frac{5}{6}. 
\end{align} 
Despite the ordering of the history for the classical protocol being different in these two cases, the quantum calculation given by definition 5 is the same for both (as it only depends on the flags raised and results obtained before Charlie measures). This leads to interference between the two causal orders of $A$ and $B$ in $\ket{\phi_{(l_3=\text{Charlie},3,H_2)}}$.

This alternative classical procedure which emulates the quantum experiment therefore gives $p^\text{ac}(000|011)=\frac{1}{2}\times\frac{1}{2}\times\frac{5}{6}+\frac{1}{2}\times\frac{1}{2}\times\frac{1}{2}\times\frac{5}{6}=5/16$ as desired. Although we have focused on only one probability here, the same method can be used to generate a full classical strategy which replicates the quantum experiment for all input and output choices.

\end{document}